\documentclass[twocolumn,prl,superscriptaddress,bibnotes]{revtex4-1}
\usepackage{color}
\usepackage{hyperref}
\usepackage{ulem}
\definecolor{red}{rgb}{0.75,0,0}
\definecolor{blue}{rgb}{0,0,0.75}
\definecolor{green}{rgb}{0,0.5,0}
\usepackage{amsmath}
\usepackage{amssymb}
\usepackage{lipsum}
\usepackage{bm}
\usepackage{xcolor}
\usepackage{graphicx}
\usepackage{verbatim}
\usepackage[T1]{fontenc}

\def\be{\begin{equation}}
\def\ee{\end{equation}}
\def\bea{\begin{eqnarray}}
\def\eea{\end{eqnarray}}

\def\besub{\begin{subequations}}
\def\eesub{\end{subequations}}

\def\bwd{\begin{widetext}}
\def\ewd{\end{widetext}}

\definecolor{ao(english)}{rgb}{0.0, 0.5, 0.0}
\definecolor{armygreen}{rgb}{0.29, 0.33, 0.13}
\definecolor{auburn}{rgb}{0.43, 0.21, 0.1}
\definecolor{brightmaroon}{rgb}{0.76, 0.13, 0.28}
\definecolor{cadmiumred}{rgb}{0.89, 0.0, 0.13}
\definecolor{carnelian}{rgb}{0.7, 0.11, 0.11}
\definecolor{cornellred}{rgb}{0.7, 0.11, 0.11}
\definecolor{crimsonglory}{rgb}{0.75, 0.0, 0.2}
\definecolor{orangeyellow}{rgb}{0.3, 0.2, 0.2}
\definecolor{fluorescentorange}{rgb}{1.0, 0.75, 0.0}
\definecolor{gamboge}{rgb}{0.89, 0.61, 0.06}
\newcommand{\bsf}[1]{\textsf{\textbf{#1}}}

\newcommand{\AM}[1]{\textcolor{black}{#1}}

\newcommand{\AMN}[1]{\textcolor{black}{#1}}
\newcommand{\SR}[1]{\textcolor{black}{#1}}

\newcommand{\AMSR}[1]{\textcolor{black}{#1}}

\begin{document}
\title{{\SR{Swimmer suspensions on substrates: anomalous stability and long-range order}}}
\author{Ananyo Maitra}
\email{ananyo.maitra@u-psud.fr}
\affiliation{LPTMS, CNRS, Univ. Paris-Sud, Universit\'e Paris-Saclay, 91405 Orsay, France}
\author{Pragya Srivastava}
\affiliation{The Francis Crick Institute, Lincoln's Inn Fields Laboratory, 44 Lincoln's Inn Fields,
London WC2A 3LY}
\author{M. Cristina Marchetti}
\affiliation{Physics Department and Syracuse Soft Matter Program, Syracuse University, Syracuse,
NY 13244, USA}
\author{Sriram Ramaswamy}
\affiliation{Centre for Condensed Matter Theory, Department of Physics, Indian Institute of Science, Bangalore 560 012, India}
\author{Martin Lenz}
\email{martin.lenz@u-psud.fr}
\affiliation{LPTMS, CNRS, Univ. Paris-Sud, Universit\'e Paris-Saclay, 91405 Orsay, France}
\begin{abstract}
\SR{We present a comprehensive theory of the dynamics and fluctuations of a two-dimensional suspension of polar active particles in an incompressible fluid confined to a substrate. {We show that, depending on the sign of a single parameter, a state with polar orientational order is \SR{anomalously stable (or anomalously unstable)},
with a nonzero relaxation (or growth) rate for angular fluctuations at zero wavenumber. This screening of the broken-symmetry mode in \AMSR{the stable} state does lead to conventional rather than giant number fluctuations as argued by Bricard et al., Nature \textbf{503}, 95 (2013), but their bend instability in a splay-stable flock does not exist \AMSR{and the polar phase has long-range order in two dimensions}.} Our theory also describes confined three-dimensional thin-film suspensions of active polar particles as well as dense compressible active polar rods, and predicts a flocking transition without a banding instability
}
\end{abstract}

\maketitle

\normalem
Biological systems are powered by energy supplied directly and independently at the level of constituent particles, and move systematically while dissipating it. This can lead to macroscopic stresses and currents responsible for diverse phenomena ranging from crawling of confluent cellular cultures to coherent motion of animal flocks. Artificial analogues of coherently moving biological systems have also been developed, most notably polar flocks in \SR{vibrated granular layers} \cite{Harsh} and self-propelled rollers \cite{Bartolo1}. ``Active hydrodynamics'' \cite{RMP, Sriram_rev, Prost_nat, Ramaswamy_Toner_Tu, Curie_rep}, which seeks to understand how nonequilibrium currents and forces affect the orientational order of anisotropic units, presents a general framework to study the large scale dynamics in such microscopically driven systems.

Active phases \SR{frequently defy expectations rooted in equilibrium physics. Motile $XY$ spins \cite{Toner_Tu, RMP, Ramaswamy_Toner_Tu, Sriram_rev} on a substrate display long-range orientational order even in two dimensions, and anomalous number fluctuations with the standard deviation in number $N$ of particles in a region growing more rapidly than $\sqrt{N}$} \cite{Ano_apol, Aditi2, Toner_Tu, RMP}. Enhancing the noise in this system leads to a disordered phase through a transition that is again distinct from its equilibrium analogue: an instability towards an inhomogeneous banded phase generically occurs between the homogeneously ordered and disordered phases ultimately rendering the transition discontinuous \cite{bertin, chate}. 

\SR{Much of our understanding of polar active systems~\cite{Toner_Tu, bertin, chate} comes from studies that ignore any ambient solvent, but biological systems are typically suspensions in an incompressible fluid that mediates long-range hydrodynamic interactions. This aspect is well dealt with for \textit{bulk} suspensions \cite{Aditi1,RMP}, but subtleties arise for a system confined to two dimensions by walls or adsorption on a substrate. The Stokesian hydrodynamic interaction, although screened at leading order by the bounding surfaces, leaks through in a weakened form through the inescapable nonlocal constraint of incompressibility \cite{Liron&Mochon, SRsedreview,Bartolo2}. }
	
\SR{In this Letter we present a general theory of polar active particles with a nonconserved velocity field on a two-dimensional substrate, taking the effects of confined incompressible flow correctly into account. In such systems motility drives flow, and flow reorients motility. These tendencies, if mutually reinforcing and strong enough, are known \cite{Harsh,Bartolo1} to lead to spontaneous flocking. }

\SR{Here are our main results. (i) Through the interplay of motility and incompressibility, such a flock is stable for \textit{all} directions of wavevector, with \textit{both bend and splay} deformations of the orientational broken-symmetry variable in such a flock relaxing on a finite, non-hydrodynamic time scale as the wavenumber $q \to 0$. This contradicts the claimed generic bend instability \cite{Bartolo1} of confined incompressible flocks, and is of course contrary to conventional expectations~\cite{MPP} of a vanishing relaxation rate in the long-wavelength limit.
(ii) \AMSR{Motility and incompressibility also suppress the instability towards the inhomogenoeus banded state that generically occurs in \emph{compressible} polar systems between the ordered and the disordered states implying that a direct transition from an isotropic state to a homogeneous flock is possible, without the intervention of a banded phase.} 
(iii) The variance of orientational fluctuations is non-divergent for $q\to 0$, with a correlation length that is finite for any nonzero motility. As a consequence number fluctuations are normal, with variance proportional to the mean \cite{Bar_footnote}.
(iv) Our main results remain correct up to very large length scales even in \AMN{weakly} compressible systems~\cite{Toner_Tu,Harsh}. Our theory is relevant to \SR{all current experiments on planar confined active polar suspensions} \cite{Bausch, Butt, Kaiser}, which we illustrate by showing how it emerges from the averaging of the dynamics of \AMN{three-dimensional polar fluid} \AMN{confined in one direction}.}

We start by constructing the general dynamical equations for the polarisation ${\bf p}({\bf r}, t)$ \AMSR{and the} concentration $c({\bf r}, t)$ \AMSR{of a collection of active units suspended in a fluid with the total velocity field of the particles and the fluid being}  ${\bf u}({\bf r}, t)$, where ${\bf r}$ is a two-dimensional position vector. \AMSR{The joint density $\rho$ of the particles and the fluid is incompressible i.e. $\dot{\rho}=0$ implying $\nabla\cdot{\bf u}=0$.} In the absence of activity and fluid flow, their equilibrium relaxation derives from a Landau-de Gennes free energy, which we write in the single Frank constant approximation for simplicity \cite{deGen}: 
\begin{equation}
\mathcal{H}=\int_{\bf r}\left[{\alpha(c)\over2}|{\bf p}|^2+{\beta\over4}|{\bf p}|^4+ {K\over2}|\nabla{\bf p}|^2+{\gamma} {\bf p}\cdot\nabla c+c\ln c\right],
\end{equation}
where the sign of $\alpha(c)$ determines the stability of the isotropic, flow-less phase. A negative $\alpha$ thus gives rise to a non-zero polarization, which remains bounded due to the confining $\beta$-term and whose heterogeneities are suppressed by the elasticity constant $K$. The $\gamma$-term describes the tendency of the polarity to align along or opposite to concentration gradients as allowed by symmetry \cite{kung}, while the last term is characteristic of an ideal solution (setting $k_BT=1$).

To lowest order in gradients, the \SR{generic} dynamical equation for ${\bf p}$ is
\begin{equation}
\label{pol}
\partial_t {\bf p}=\Lambda {\bf u}-\frac{\delta \mathcal{H}}{\delta {\bf p}},
\end{equation}
where the coefficient $\Lambda$ aligns the polarisation vector with the local \SR{suspension} velocity and is specific to systems in contact with a substrate~\cite{Harsh, Brotto, Dadhichi_JSTAT}.
We treat it as independent \SR{of} the direction of ${\bf p}$, which does not qualitatively modify our conclusions~\cite{supp}. The coefficient in front of the second term of the right-hand side of Eq.~\eqref{pol} is set to one through a proper choice of time units.
Similarly, \SR{ignoring inertia, Newton's second Law reduces to force balance which} to lowest order in gradients is
\begin{equation}
\label{vel}
\Gamma{\bf u}=\upsilon{\bf p}-\nabla\Pi-\Lambda\frac{\delta \mathcal{H}}{\delta {\bf p}},
\end{equation}
where $\Gamma$ \SR{is the coefficient of damping by} the substrate and $\Pi$ is the pressure that enforces incompressibility, \SR{\textit{i.e.}}, $\nabla\cdot{\bf u}=0$. The $\upsilon {\bf p}$ denotes the active \AMSR{polar force density} of the particles. The final term on the right-hand-side of Eq. \eqref{vel} is required by Onsager symmetry to ensure that the steady-state in the limit of vanishing activity reduces to the equilibrium distribution. Finally, the continuity equation for the concentration field \AMSR{to second order in gradients} is
\AMSR{\begin{equation}
\label{conc}
\partial_tc+{\bf u}\cdot\nabla c=-\nabla\cdot\SR{\left(\upsilon_pc{\bf p}-D_c c\nabla \frac{\delta \mathcal{H}}{\delta c}\right)}.
\end{equation}}
where \AM{$v_pc{\bf p}$ }denotes an active \AMSR{concentration} current due to the motility of the particles \cite{Toner_Tu, RMP, Toner_rean, Ramaswamy_Toner_Tu, Sriram_rev}. The second term of its right-hand-side leads to standard diffusive dynamics. Equations (\ref{pol}-\ref{conc}) are similar to the ones for a two-fluid polar active system \cite{Harsh}, with the only difference being the incompressibility \AMSR{of the joint density of the particles and the fluid}.

{To determine the stability of a homogeneous isotropic ($|{\bf p}|=0$) state, we perform a linear stability analysis of Eqs. \eqref{pol} and \eqref{vel}. For $\Lambda\upsilon>0$, the state is destabilised when \SR{\cite{Harsh}}
\begin{equation}
\label{disord}
\tilde{\alpha}=\alpha(c)-\frac{\Lambda\upsilon}{\Gamma+\Lambda^2}=\alpha(c)-w<0.
\end{equation}
Thus, for \AM{$\upsilon>0$, a positive alignment parameter $\Lambda$} favours the instability of the homogeneous disordered phase, as it reinforces the particles' alignment once they have started moving~\cite{Brotto}. Following this instability, a homogeneous ordered phase with ${\bf p}=p_0\hat{x}$, $c=c_0$ and ${\bf u}=u_0\hat{x}$ may form, where $p_0^2=|\tilde{\alpha}/\beta|$ and $u_0=(w/\Lambda)p_0$. To study its stability, we project Eq. \eqref{vel} transverse to the wavevector ${\bf q}$ to eliminate the velocity field. Introducing the polarisation fluctuations $\delta {\bf p}=(p_0+\delta p)(\cos\theta\hat{x}+\sin\theta\hat{y})-p_0\hat{x}$, we find closed form equations for small deviations from the ordered state: $\partial_t(\delta c, \delta p, \theta)={\bsf M}\cdot(\delta c, \delta p, \theta)$. The three eigenvalues of matrix ${\bsf M}$ each characterize a relaxation mode of the system. Naively, the presence of both a conservation law for the concentration and of a broken continuous symmetry in the orientation of ${\bf p}$ would suggest that two of these modes should relax on slow `hydrodynamic' time scales, \SR{\textit{i.e.}}, that the associated eigenvalues vanish in the ${\bf q}\to 0$ limit. A detailed calculation however reveals that this is not correct~\cite{supp} in the presence of the long-range hydrodynamic  interactions mediated by the incompressibility of the fluid.
Such long-range interactions typically suppress fluctuations in the ordered state and thus stabilize it, as is the case in dipolar XY models \cite{Maleev}. Here, they imply that our system has only one hydrodynamic mode {{associated with the relaxation of the concentration}} with relaxation rate $\kappa_c\propto q^2$, the stability of which we discuss later. 
The remaining two eigenvalues, which govern the dynamics of the polarization fluctuations, are non-hydrodynamic, and thus go to a finite limit as ${\bf q}\to 0$, namely
\begin{align}
\label{eigen}
\kappa_\pm=&-\frac{1}{2}\left[w+2|\tilde{\alpha}|\left(1+\frac{\Lambda^2}{\Gamma}\sin^2\phi\right)\right.\\
&\left.\pm\sqrt{\left[w+2|\tilde{\alpha}|\left(1+\frac{\Lambda^2}{\Gamma}\sin^2\phi\right)\right]^2-8|\tilde{\alpha}|w\sin^2\phi\left(1+\frac{\Lambda^2}{\Gamma}\right)}\right],\nonumber
\end{align}
where $\phi$ is the angle between ${\bf q}$ and $\bf\hat{x}$. For $w<0$, one of the eigenvalues is always positive, implying a generic instability. For $w>0$, both $\kappa_+$ and $\kappa_-$ are stabilising, though one of the two eigenvalues vanish for fluctuations whose wavevectors are aligned precisely in direction of ordering. This implies that all components of the polarisation vector \SR{have a finite exponential-decay time to their steady state value} even in infinite systems. The ordered state is thus \SR{exceptionally stable} to polarisation fluctuations, \SR{a consequence of both} activity-induced motility \AMN{$\upsilon$} and long-range hydrodynamic interactions \AMN{due to incompressibility}.

While Eq.~\eqref{eigen} demonstrates that two of the eigenvalues of the dynamical matrix ${\bsf M}$ are stabilising, its third and only hydrodynamic eigenvalue $\kappa_c(\phi)$ also has to be negative to ensure that a homogeneous polar phase exists (see the supplemental material for the full expression of $\kappa_c$~\cite{supp}). 
However, close to the transition (\emph{i.e.}, for $\tilde{\alpha}\rightarrow 0^-$), this eigenvalue is known to always turn positive for $\phi=0$ in compressible systems \cite{RMP}, implying a generic instability towards a non-homogeneous, banded phase~\cite{bertin, chate, Solon2, Solon3}. In our incompressible system however, this instability \AMSR{is} suppressed \AMSR{for all $-\gamma\upsilon_p<D_cw/c_0$, which is simply the condition for the stability of the homogneous flock}:
\AMSR{\begin{equation}
\label{band}
\lim_{\tilde{\alpha}\rightarrow 0}\kappa_c(0)=
-\left[D_c+\frac{c_0\gamma\upsilon_p}{w}\right]q_x^2
\end{equation}}
As a result, the \AMSR{transition to the ordered state} in a \AMSR{polar suspension in an incompressible fluid} \AMSR{will} not \SR{necessarily} proceed \AMSR{via} a \AMSR{banded phase} and can thus differ from that of \AMSR{flocks in the absence of a fluid medium}. 

To determine the effect of noise on the ordered phase of our system, we first compute the static structure factor of its angular fluctuations in the presence of an additional zero-mean Gaussian white noise $\xi({\bf r}, t)$ in the right-hand side of Eq.~\eqref{vel} with $\langle\xi({\bf r}, t)\xi({\bf r'}, t')\rangle=2B\delta({\bf r}-{\bf r'})\delta(t-t')$. In the aligned phase, this yields:}
\begin{equation}
\label{corr}
S(q,\phi)=\langle|\theta({\bf q})|^2\rangle=\frac{B}{K_p(\phi)q^2+w \sin^2\phi}
\end{equation}
where the somewhat cumbersome form of $K_p(\phi)$ is given in the supplement~\cite{supp}. The integral of this structure factor over wavevectors $\bf q$ converges, implying a finite amplitude for the angular fluctuations and thus the existence of a long-range ordered aligned polar phase.
To verify that this conclusion, which we obtained by linearizing Eqs.~(\ref{pol}-\ref{conc}), is not modified by the inclusion of nonlinearities, we consider the simple case of a 
\SR{flock in which number is not conserved} \cite{Malthusian}.
In this simple case, our model exactly maps onto the \SR{polar flock} with constraint $\nabla\cdot {\bf p}=0$ studied in Ref.~\cite{CLT}. This mapping ultimately yields exact equal-time exponents of the ordered phase via a transformation to the KPZ equation~\cite{supp}. This implies that our model, in which there is no explicit constraint on ${\bf p}$, also has long-range order in two dimensions with the same roughness and anisotropy exponents as in Ref.~\cite{CLT}. This relation between the nonlinear theory of \SR{polar swimmers without number conservation} in incompressible polar fluid and a theory of a \SR{suspension of} polar active particles with $\nabla\cdot{\bf p}=0$ is unusual; for instance, an \SR{\textit{apolar}} system in an incompressible fluid described in \cite{Ano_apol} does not correspond to a theory in which $\nabla\nabla:{\bsf Q}=0$, where ${\bsf Q}$ is the apolar order parameter. In addition, removing the condition of fixed concentration introduces additional relevant nonlinearities and spoils the mapping, likely resulting in an ordered phase with distinct behaviour.

Beyond the existence and stability of an ordered phase in two dimensions, a hallmark of active matter physics is the possibility of much larger density fluctuations than allowed by equilibrium physics. To assess their existence in our system, we add a non-conserving Gaussian white noise to Eq.~\eqref{pol} and a conserving noise to Eq.~\eqref{conc}. We find that the number fluctuations now scale as $\sqrt{N}$ as in equilibrium systems \cite{supp}, despite the presence of an active particle current $\propto {\bf p}$ in Eq.~\eqref{conc}. 
Indeed, since all components of ${\bf p}$ have fast, non-hydrodynamic relaxation rates, the polarisation quickly aligns with any gradient in concentration and ${\bf p}\sim\nabla c$ on long (hydrodynamic) time and length scales. As a result, the active $\propto {\bf p}$ current is equivalent to a passive diffusive current, implying equilibrium-like statistics for the concentration fluctuations. This is a direct result of \SR{the nonzero restoring torque of our system for orientational distortions even in the limit of long wavelengths} \cite{Bar_footnote}.

While the above results are directly relevant for the experiments on single layers of motile particles \cite{Bartolo1}, our theory also describes the effective thickness-averaged dynamics of three-dimensional films of polar active particles in an incompressible fluid of lateral dimension $L$, confined along the $z$ direction over a length scale $h\ll L$. 
To demonstrate this, we describe the three-dimensional polar fluid by the three-dimensional polarisation vector field $\bar{{\bf p}}(\bar{\bf x},t)=(\bar{{\bf p}}_\perp, \bar{p}_z)$, velocity $\bar{{\bf u}}(\bar{\bf x},t)=(\bar{{\bf u}}_\perp, \bar{u}_z)$ and particle number $\bar{c}(\bar{\bf x},t)$, where $\bar{\bf x}$ is a three-dimensional position vector, and $\bar{{\bf p}}_\perp$ and $\bar{p}_z$ and $\bar{{\bf u}}_\perp$ and $\bar{u}_z$ are the projections transverse to and along the confining direction of the three-dimensional  polarisation and velocity respectively. We further denote the three-dimensional gradient by $\bar{\nabla}$. We describe the three-dimensional dynamics of our system using a standard set of constitutive equations~\cite{Ano_membrane, Cristina_polar}, and describe our detailed thickness-averaging calculation in the supplementary material~\cite{supp}. Essentially, we use the lubrication approximation of thin-film flows \cite{Stone} to project our equations in two dimensions,  exploiting the fact that the gradients along $z$ are large, namely $\partial_{\bar{z}}={\cal{O}}(1/h)\gg \partial_{\bar{x}},{\partial_{\bar{y}}}$. \AM{As is standard in such theories,} the thickness average of the three-dimensional viscous force density $\bar{\eta}\bar{{\nabla}}^2\bar{{\bf u}}$, with $\bar{\eta}$ being the viscosity of the three-dimensional fluid, yields the friction-like force  $-\Gamma{\bf u}$ in Eq.~\eqref{vel} to lowest order in $h/L$, where $\Gamma=12\bar{\eta}/h^2$ and ${\bf u}$ is the thickness-averaged velocity in the $xy$ plane, with the three-dimensional incompressibility condition translating into $\nabla\cdot{\bf u}=0$.
Beyond this standard viscous force and other passive terms reminiscent of classical hydrodynamics, our three-dimensional dynamical equations feature two three-dimensional active polar force densities, namely $\bar{\nabla}^2\bar{{\bf p}}$ and $\bar{\nabla}\cdot(\bar{{\bf p}}\bar{{\bf p}})$~\cite{Cristina_polar}.
The former characterises the fore-aft symmetry around, and hence the motility of, an elementary active object. Upon thickness averaging it leads to the two-dimensional propulsive force $\propto {\bf p}$ of Eq.~\eqref{vel}, where ${\bf p}$ is the thickness average of the transverse polarization $\bar{{\bf p}}_\perp$. The latter active term determines the contractile or extensile character of active units, and leads to a force $\propto \nabla\cdot({\bf pp})$, which is subdominant at large lateral scales and is thus not included in our two-dimensional equations.
We similarly obtain Eq.~\eqref{pol} for the polarisation field by choosing walls forcing a non-trivial $\bar{z}$-dependence on the polarisation $\bar{{\bf p}}$ through the boundary conditions $\bar{{\bf p}}_{\bar{z}=0}=\hat{z}$ and $\bar{{\bf p}}_{\bar{z}=h}=-\hat{z}$. Polarization is generically affected by shear, which we describe through the symmetric strain rate tensor $\bar{{\bsf U}}=(1/2)[\bar{\nabla}\bar{\bf u}+(\bar{\nabla}\bar{\bf u})^T]$. This coupling gives rise to two different contributions to $\partial_t\bar{\bf p}$ in the three-dimensional polarization equation, namely $\bar{\nabla}\cdot\bar{{\bsf U}}=(\bar{\nabla}\cdot\bar{\nabla})\bar{{\bf u}}$ which describes the alignment of the polarisation vector with with the local \emph{gradients} of the shear rate, and $\bar{{\bf p}}\cdot\bar{{\bsf U}}$, which describes its alignment to a local shear flow. Again using lubrication arguments, we obtain the first term on the right-hand-side of Eq. \eqref{pol} from the former. The latter leads to the usual flow-alignment, which aligns the polarity with the two-dimensional velocity \emph{gradient} and due to its subdominance is not included in \eqref{pol}. Finally, the thickness-averaged concentration equation \eqref{conc} is obtained by imposing no-flux boundary condition on the three-dimensional continuity equation for $\bar{c}$.

\begin{figure}
  \centering
  \includegraphics[width=6cm]{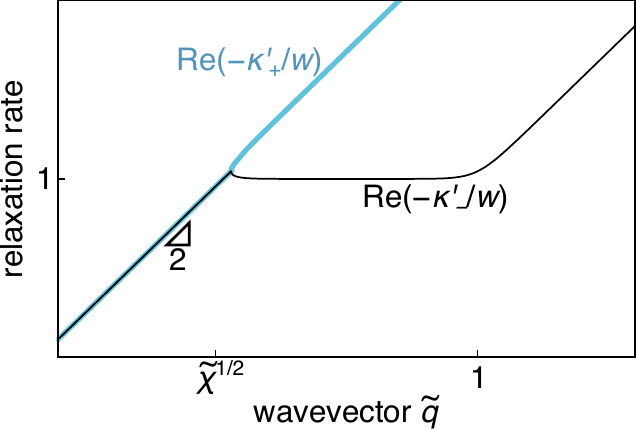}
\caption{\AMN{A log-log plot of the dimensionless} decay rates for the two eigenmodes associated with the coupled dynamics of the total density and angular fluctuations in a weakly compressible system [Eq.~\eqref{eq:CompressibleRates}]. While both modes display a diffusive (slope 2) relaxation at low dimensionless wavevector $\tilde{q}$, taking the dimensionless compressibility $\tilde{\chi}$ to zero shifts the blue curve to the left, implying that the relaxation rate associated with any finite $\tilde{q}$ goes to infinity. Meanwhile, the second relaxation rate (black curve) develops a wide $\tilde{q}$-independent plateau, mimicking the non-hydrodynamic relaxation rate associated with a truly incompressible system.}
\label{fig1}
\end{figure}

While strictly valid for incompressible systems, our conclusions regarding \AMSR{the non-hydrodynamic relaxation of angular fluctuations} and non-giant \AMSR{number} fluctuations are also applicable up to very large length scales in weakly compressible systems such as fluid-less collections of motile particles or active polar rods in a dense bead medium~\cite{Harsh}. To characterize such systems, we \AMSR{reintroduce} the dynamics of their overall density field $\rho$ 
\AMSR{which  now} satisfies the conservation equation $\partial_t\rho=-\nabla\cdot(\rho{\bf u})$. We disregard the relaxation of the polar particle concentration $c$ for simplicity, and consider only the coupled dynamics of $\rho$ and of small fluctuations in the angle $\theta$ described by Eqs.~(\ref{pol}-\ref{vel}). We assume a system deep in the ordered phase, implying that $\delta p$ quickly relaxes to zero on hydrodynamic time scales. \SR{We assume a linear relation between small changes in the pressure $\Pi$ and the density $\rho$: $\Pi(\rho) - \Pi(\rho_0) \simeq \delta\rho/(\chi\rho_0)$}, where $\chi$ is the fluid's compressibility, $\rho_0$ is the average density and $\delta\rho$ is the deviation of the local density from its average value. Defining the \SR{non-dimensional} compressibility $\tilde{\chi}=\chi K_p\rho_0\Gamma$, we first \AMSR{check the fate of the relaxation rate of angular fluctuations  (the wavevector-independent relaxation rate being strictly valid only for $\tilde{\chi}=0$)} in our weakly compressible fluid ($\tilde{\chi}\ll 1$). We focus our discussion on the direction $\phi=\pi/2$, which displays the strongest incompressibility-induced stabilization in the incompressible case. The eigenvalues associated with the coupled relaxation of the density and orientational fluctuations $(\delta\rho,\theta)$ read
\begin{equation}\label{eq:CompressibleRates}
\kappa'_\pm=-\frac{w\tilde{q}^2}{2\tilde{\chi}}\left[1+\tilde{\chi}\pm
\sqrt{(1-\tilde{\chi})^2-\frac{4\mathcal{\tilde{\chi}}}{\tilde{q}^2}}\right].
\end{equation}
with $\tilde{q}=q\sqrt{K/w}$. We see in Fig.~\ref{fig1} that $\kappa'_+\sim -w\tilde{q}^2/\tilde{\chi}$ becomes infinitely large as $\tilde{\chi}\rightarrow 0$, indicating that the pressure homogenises very quickly in a nearly-incompressible medium. Following this rapid homogenisation, the alignment direction $\theta$ relaxes at a rate $\kappa'_-$. In an incompressible system, this relaxation rate was non-hydrodynamic, \emph{i.e.}, went to a finite limit as $q\rightarrow 0$, leading to \AMSR{polar-ordered states that were singularly less susceptible to angular fluctuations than any equilibrium system or even \emph{highly compressible} flocks \cite{Toner_Tu}}. This is not strictly the case here, as $\kappa'_- \propto -\tilde{q}^2$ for very small wavevectors \AMN{$\tilde{q}<{\tilde{\chi}}^{1/2}$}. However, $\kappa'_-$ has a plateau for intermediate wavevectors \AMN{${\tilde{\chi}}^{1/2}<\tilde{q}<1$} that \AMN{extends to $q\to 0$} for $\tilde{\chi}\rightarrow 0$. As \AMN{the smallest wavevector realisable in a system of size $L$ is \SR{$\pi/L$}, this implies that a weakly compressible polar fluid will be indistinguishable from a truly incompressible polar system as long as $L\ll 1/\sqrt{\chi w \rho_0\Gamma}$.}


Furthermore, to assess the effect of noise, we recompute the structure factor of Eq.~\eqref{corr} for our weakly compressible system to find
\begin{equation}
\label{ssff}
S^{\tilde{\chi}}(q,\pi/2)=\left(1+\frac{\tilde{\chi}}{1+\tilde{\chi}}\tilde{q}^{-2}\right)S(q,\pi/2),
\end{equation}
where $S(q)$ is the incompressible structure factor of Eq.~\eqref{corr}. Again, we find that our weakly compressible system behaves as an incompressible one down to \AMN{$\tilde{q}\approx \sqrt{\tilde{\chi}}$}, implying that intermediate-size compressible systems are deprived of giant-number fluctuations similar to strictly incompressible systems, while very large ones do display them.

While our discussion has so far focused on polar particles that align with the local flow ($w>0$), consistent with existing experiments~\cite{Harsh, Bartolo1}, systems with $w<0$ are \SR{conceivable \cite{GuptaGupta}}. One possibility would be particles that point opposite to the local flow ($\Lambda<0$) while moving along their polarity ($\upsilon>0$). In this case the homogeneous ordered phase is unstable and all perturbations with wavenumber smaller than $\sqrt{K/(|w|\sin^2\phi)}$ grow exponentially. 

The analysis presented here clarifies theoretical expectations on the structure of number fluctuations in existing experiments on active systems in an incompressible fluid, which in conjunction with experimental challenges have been a source of confusion~\cite{Bartolo1, Bartolo2}.
It moreover provides a framework to analyze the dynamics of numerous quasi-2D biological systems, which are \SR{almost invariably} immersed in an incompressible fluid, from the scale of the intracellular medium~\cite{Bausch} to that of crawling cell layers~\cite{ccl}.
Its predictions of non-hydrodynamic relaxation, the possible absence of a banding phase at the disorder-order transition and normal number fluctuations should be testable in any of these contexts or in artificial chemotactic colloids~\cite{saha}; such artificial systems could additionally be engineered to align antiparallel to the motility direction ($\Lambda\upsilon<0$), which we predict will abolish long-range order altogether. Our results are also largely applicable to weakly compressible systems such as dense granular layer of polar rods or dense mixtures of rods and beads \cite{Harsh}.
From a theoretical standpoint, our work establishes that \SR{the hydrodynamic interaction \AMSR{singularly} alters {\it equal-time} as well as time-displaced correlations of the orientation \AMSR{even when the long-wavelength fluctuations of the fluid momentum-density are \SR{damped by friction with a substrate}}.  
Alongside the now-classic breaking of the Hohenberg-Mermin-Wagner theorem~\cite{Toner_Tu, Ramaswamy_Toner_Tu} and existence of anomalously large fluctuations~\cite{Toner_Tu, RMP, Sriram_rev}, this finding constitutes another striking violation of equilibrium expectations in active matter.} 

\begin{acknowledgments}
ML was supported by Marie 
Curie Integration Grant PCIG12-GA-2012-334053, ‘‘Investissements 
d’Avenir’’ LabEx PALM (ANR-10-LABX-0039-PALM), ANR grant 
ANR-15-CE13-0004-03 and ERC Starting Grant 677532. ML’s group belongs to 
the CNRS consortium CellTiss. SR was supported by a J C Bose Fellowship of the SERB (India) and the Tata Education and Development Trust. SR also thanks the Simons Foundation and the KITP for support. MCM was supported by the National Science Foundation  through award DMR-1609208.
\end{acknowledgments}

\end{document}